\documentclass[10pt, twocolumn, pre, aps, superscriptaddress, showpacs]{revtex4-1}
\usepackage{amsmath, graphicx, subfigure, tikz}
\begin{document}

\title{First-Order to Second-Order Phase Transition Changeover and Latent Heats\\
of q-State Potts Models in d=2,3 from a Simple Migdal-Kadanoff Adaptation}

\author{H. Ya\u{g}{\i}z Devre}
     \affiliation{\"Usk\"udar American Academy, \"Usk\"udar, Istanbul 34664, Turkey}
\author{A. Nihat Berker}
    \affiliation{Faculty of Engineering and Natural Sciences, Kadir Has University, Cibali, Istanbul 34083, Turkey}
    \affiliation{T\"UBITAK Research Institute for Basic Sciences, Gebze, Kocaeli 41470, Turkey}
    \affiliation{Department of Physics, Massachusetts Institute of Technology, Cambridge, Massachusetts 02139, USA}

\begin{abstract}
The changeover from first-order to second-order phase transitions in q-state Potts models is obtained at $q_c=2$ in spatial dimension $d=3$ and essentially at $q_c=4$ in $d=2$, using a physically intuited simple adaptation of the Migdal-Kadanoff renormalization-group transformation.  This simple procedure yields the latent heats at the first-order phase transitions.  In both $d=2$ and 3, the calculated phase transition temperatures, respectively compared with the exact self-duality and Monte Carlo results, are dramatically improved.
\end{abstract}
\maketitle

\section{Introduction: Order of Potts Transitions and Underlying Physical Intuition}

The spatial dimensionality $d$, the symmetry of the local degrees of freedom, and the presence of quenched randomness strongly affect the occurrence and order of a phase transition.  A simple but effective method in studying the occurrence of a phase transition has been the renormalization-group method under the Migdal-Kadanoff approximation \cite{Migdal,Kadanoff}, which is also currently the most used renormalization-group method.  Thus, using this method on widely different systems, the lower-critical dimension $d_c$ below which no ordering occurs has been correctly determined as $d_c=1$ for the Ising model \cite{Migdal,Kadanoff}, $d_c=2$ for the XY \cite{Jose,BerkerNelson} and Heisenberg \cite{Tunca} models, and the presence of an algebraically ordered phase has been seen for the XY model \cite{Jose,BerkerNelson,Sariyer}.  In systems with frozen microscopic disorder (quenched randomness), using the simple Migdal-Kadanoff renormalization-group approximation, $d_c=2$ has been determined for the random-field Ising \cite{Machta,Falicov} and XY models \cite{Kutay}, and, yielding a non-integer value, $d_c=2.46$ for Ising spin-glass systems \cite{Atalay}. Also under the Migdal-Kadanoff approximation, the chaotic nature of the spin-glass phases \cite{McKayChaos,McKayChaos2,BerkerMcKay} has been obtained and quantitatively analyzed, both for quenched randomly mixed ferromagnetic-antiferromagnetic spin glasses \cite{Ilker1,Ilker2,Ilker3} and right- and left-chiral (helical) spin glasses \cite{Caglar1,Caglar2,Caglar3}.

\begin{figure}[ht!]
\centering
\includegraphics[scale=0.4]{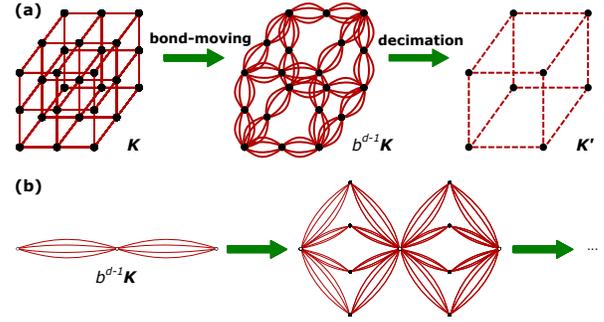}
\caption{From Ref.\cite{Artun}: (a) Migdal-Kadanoff approximate renormalization-group
transformation for the $d=3$ cubic lattice with the length-rescaling
factor of $b=2$. (b) Construction of the $d=3, b=2$ hierarchical
lattice for which the Migdal-Kadanoff recursion relation is exact.}
\end{figure}

\begin{figure*}[ht!]
\centering
\includegraphics[scale=0.33]{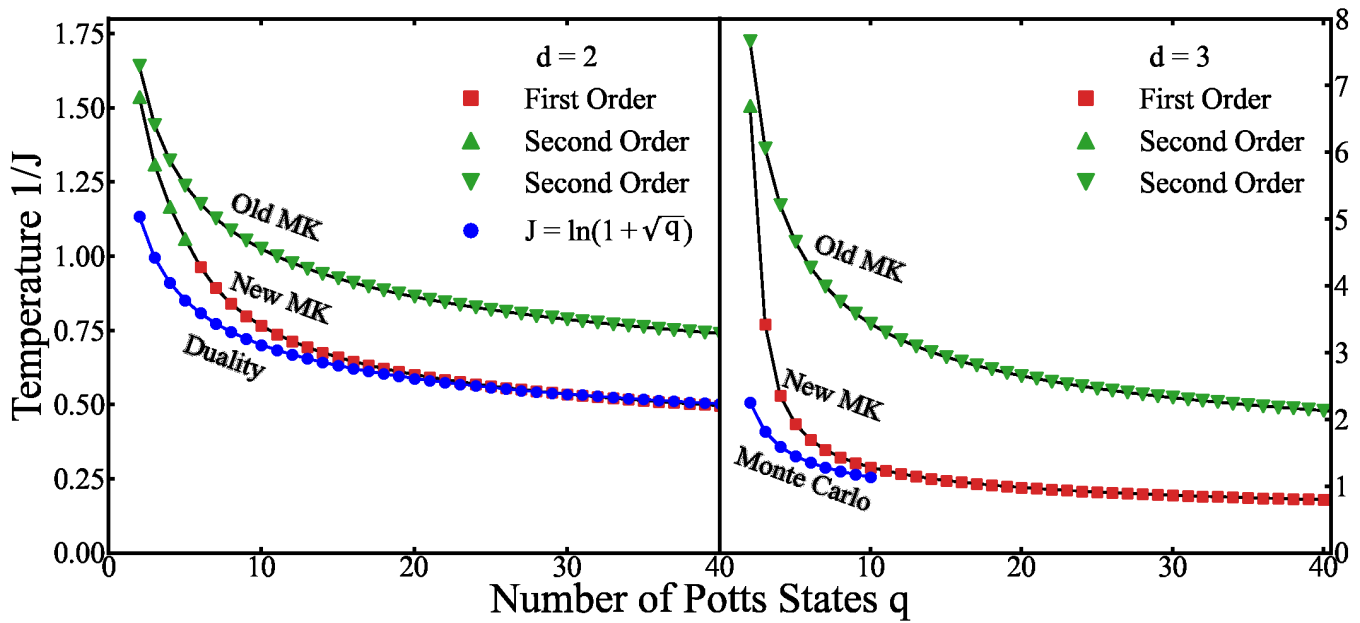}
\caption{Calculated transition temperatures $1/J$ of $q$-state Potts models.  The top curve is obtained with the conventional Migdal-Kadanoff approximation.  In $d=2$, the bottom curve is the exact transition temperatures obtained from self-duality.  In $d=3$, the bottom curve is Monte Carlo results \cite{MonteCarlo}.  The intermediate curve is obtained with our simply improved Migdal-Kadanoff approximation.  First- and second-order phase transitions are given with triangles and squares, respectively.  The improved calculation gives the changeover from second- to first-order exactly (after $q=2$) in $d=3$ and very nearly (after $q=5$ instead of $q=4$) in $d=2$.  In the latter case, the changeover can be brought down to $q=4$ by a simple physical argument and calculation, as seen in Fig.4.  Both in $d=2$ and 3, the values of the phase transition temperatures are dramatically impoved with the improved calculation and join the exact resuts for $q\gtrsim 10$ and $q\gtrsim 5$, respectively.}
\end{figure*}

An important aspect of an occurring phase transition is the order of the phase transition.  The simple Migdal-Kadanoff approximation has not been successful in predicting this for an order-disorder phase transition in a model system.  The best example are the $q$-state Potts models which, in terms of model system variety and experimental application, offer rich behaviors.  The Potts models are defined by the Hamiltonian
\begin{equation}
- \beta {\cal H} = J \sum_{\left<ij\right>} \, \delta(s_i, s_j),
\end{equation}
where $\beta=1/k_{B}T$, at lattice site $i$ the Potts spin $s_{i}=1,2,,...,q$ can
be in $q$ different states, the delta function $\delta(s_i,
s_j)=1(0)$ for $s_i=s_j (s_i\neq s_j)$, and the sum is over all
interacting pairs of spins.  The Ising model is obtained for $q=1$.  The lower-critical dimension of the Potts models is $d_c=1$, as also seen by the simple Migdal-Kadanoff renormalization-group approximation.\cite{BerkerOstlund}  However, for $d > 1$, the phase transitions of the Potts models are first order for $q > q_c$ and second order for $q < q_c$.\cite{spinS7,Nienhuis1,Nienhuis2,AndelmanPotts0,AndelmanPotts1,Nienhuis3,d1,AndelmanPotts2}  This has not been obtained by the simple Migdal-Kadanoff approximation, which yields second order for all $q$.  The actual changeover number of states $q_c(d)$ depends on dimensionality $d$.  For $d=2$ and 3, $q_c=4$ and 2, respectively.  For $d=1$, $q_c=\infty$.\cite{d1}

As noted above, the $q$-state Potts models have a second-order phase transition for $q\leq q_c$ and a first-order phase transition for $q>q_c$.  In renormalization-group theory \cite{spinS7,AndelmanPotts2}, the latter has been seen understood, and reproduced, as a condensation of effective vacancies formed by regions of disorder.  Disorder is entropically favored for high number of states $q$.  However, these renormalization-group calculations have required flows in large Hamiltonian parameter spaces, with many different types of interactions, and not connectable to the phase transition temperatures or thermodynamic properties of the original Potts models (Eq.(1)).  The effective vacancy mechanism has not been incorporated into the simple, pliable, otherwise effective, and therefore much used Migdal-Kadanoff transformation.

In this study, we find an also simple, physically motivated adjustment to the usual Migdal-Kadanoff approximation that cures the problem of the order of the phase transition, dramatically improves the calculated transition temperatures both in $d=2$ and 3, and appears to be widely applicable to other systems.

\section{Migdal-Kadanoff as a Simple Effective Renormalization Group}

The Migdal-Kadanoff approximation renders a non-doable renormalization-group transformation doable by a physically motivated approximate step, is very easily calculated, applicable to large number of systems, including for example such complexities as the quenched-random helical spin glass \cite{Caglar1,Caglar2,Caglar3}, and effective across physical dimensions $d$.

\begin{figure*}[ht!]
\centering
\includegraphics[scale=0.275]{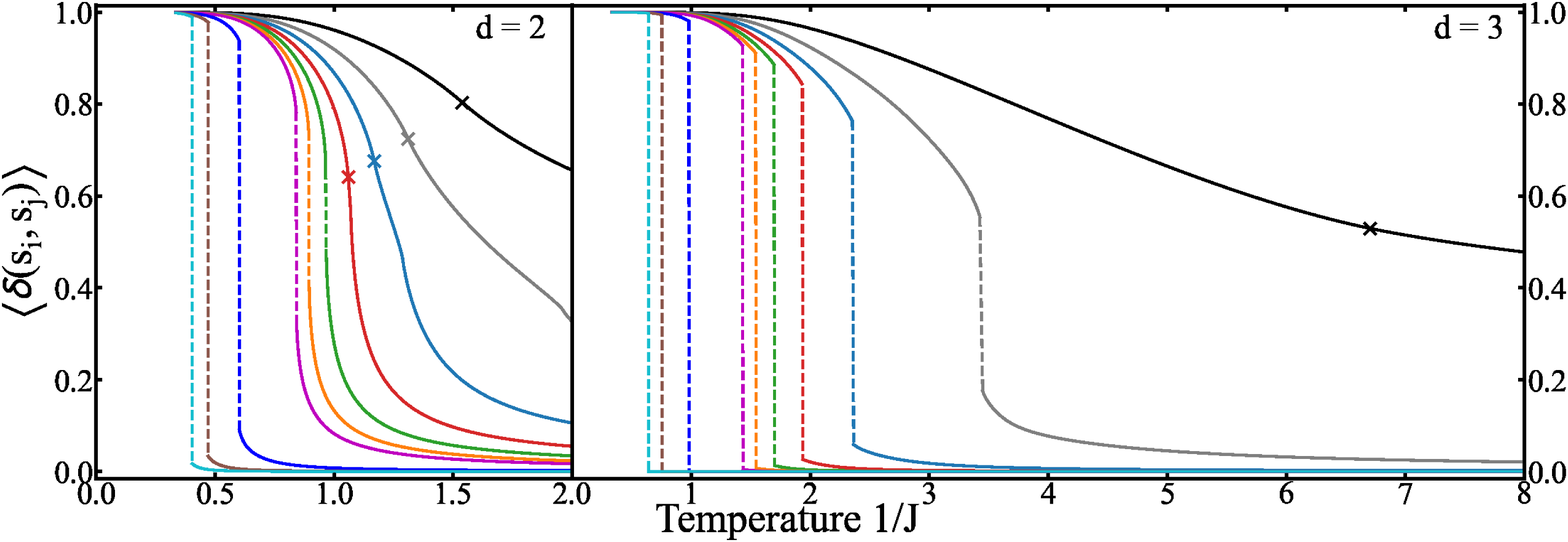}
\caption{Calculated $q$-state Potts energy densities in $d=2$ and 3.  In each panel, the curves are, from right to left, for $q=2,3,4,5,6,7,8,20,50,100$.  The latent-heat discontinuities of the first-order phase transitions are shown with the dashed lines. The second-order phase transitions are marked with x.}
\end{figure*}

Starting with the example given in Fig.(1a), an exact renormalization-group transformation cannot be applied to the cubic lattice.  Thus, as an approximation, some of the bonds are removed.  However, this weakens the connectivity of the system and, to compensate, for every bond removed, a bond is added to the remaining bonds.  This whole step is called the bond-moving step and constitutes the approximate step of the renormalization-group transformation.  At this point, the intermediate sites can be eliminated by an exact summation over their spin values in the partition function, which yields the renormalized interaction between the remaining sites.  This is called the (exact) decimation step and completes the renormalization-group transformation. As shown in Fig.1, the renormalization-group recursion relations of the Migdal-Kadanoff approximation are identical to those of an exactly solved hierarchical lattice.\cite{BerkerOstlund,Kaufman1,Kaufman2}

The above can be rendered algebraically in the most straightforward way by writing the transfer matrix between two neighboring spins,
\begin{multline}
\textbf{T}_{ij} \equiv e^{-\beta {\cal H}_{ij}} =
\left(
\begin{array}{ccc}
e^J & 1 & 1 \\
1 & e^J & 1 \\
1 & 1 & e^J \end{array} \right),
\end{multline}
where $-\beta {\cal H}_{ij}$ is the part of the Hamiltonian between the two spins at the neighboring sites $i$ and$j$.

The bond-moving step of the Migdal-Kadanoff approximate renormalization-group tranformation consists in taking the power of $b^{d-1}$ of each element in this matrix, where $b$ is the length-rescaling factor of the renormalization-group transformation, namely the renormalized nearest-neighbor separation in units of unrenormalized nearest-neighbor separation.  The decimation step consists in matrix-multiplying $b$ bond-moved transfer matrices.  The flows, under this transformation, of the transfer matrices determines the phase transition and all of the thermodynamic densities of the system, as illustrated below.

\section{Simply Improved Migdal-Kadanoff Renormalization-Group Method}

The above is cured simply by including a local disorder state into the two-spin tansfer matrix of Eq.(2).  Inside an ordered region of a given spin value, a disordered site does not contribute to the energy in Eq.(1), but has a multiplicity of $q-1$.  This is equivalent to the logarithm of an on-site energy and, with no approximation, is shared on the tranfer matrices of the $2d$ incoming bonds.  The transfer matrix does becomes
\begin{multline}
\textbf{T}_{ij} \equiv e^{-\beta {\cal H}_{ij}} = \\
\left(
\begin{array}{cccc}
e^J & 1 & 1 &(q-1)^{1/2d}\\
1 & e^J & 1 &(q-1)^{1/2d}\\
1 & 1 & e^J &(q-1)^{1/2d}\\
(q-1)^{1/2d} &(q-1)^{1/2d} &(q-1)^{1/2d} &(q-1)^{1/d}\end{array} \right).
\end{multline}

Using this transfer matrix, the renormalization-group calculation yields $q_c$.  The first-order phase transition is recognized by the disordered side at the phase transition having, under repeated recalings, the effective-vacancy position of $(q+1)\times (q+1)$ dominant in the transfer matrix, rather than the elements of the $q\times q$ upper-left submatrix being simultaneously dominant.  The first-order phase transition will be explicitly seen with the calculation, using this Migdal-Kadanoff transformation, of the latent heat

The phase transition temperatures $1/J$ of $q$-state Potts models, calculated with the simply improved Migdal- Kadanof transformation, are shown in Fig.1.  The top curve in this figure is obtained with the conventional Migdal-Kadanoff approximation.  In $d=2$, the bottom curve is the exact transition temperatures obtained from self-duality \cite{duality1}. In $d=3$, the bottom curve is Monte Carlo results \cite{MonteCarlo}.  The intermediate curve is obtained with our simply improved Migdal-Kadanoff approximation.  First- and second-order phase transitions are distinguished in the figure.  The improved calculation gives the changeover from second- to first-order exactly (after $q=2$) in $d=3$ and very nearly (after $q=5$ instead of $q=4$) in $d=2$.  In the latter case, the changeover can be brought down to $q=4$ by a simple physical argument and calculation, as seen below. Both in $d=2$ and 3, the values of the phase transition temperatures are dramatically impoved with the improved calculation and join the exact resuts for $q\gtrsim 10$ and $q\gtrsim 5$, respectively.

\section{Latent Heats of the q-State Pots Models in d=2 and 3}

The position-space renormalization-group solution of a system yields the entire statistical mechanics of the system, at and away from the phase transions, including the thermodynamic densities.\cite{Artun}  The calculation of the latter requires following the entire range of renormalization-group trajectories.  In the ordered phases, the trajectories lead to strong coupling behavior. To avoid numerical overflow problems, with no approximation, at the beginning of a trajectory and after each decimation, the transfer matrix is divided by is largest element, so that its largest element then becomes 1.  This division is equivalent to subtracting to constant from the Hamiltonian.  This division is not necessary after bond-moving, since the largest element, taken to the power $b^{d-1}$, remains 1.  The logarithm of the dividing element, namely the subtractive constant $G(n)=\ln (T_{ij})_{max}$, where $n$ indicates the $(n)$th renormalization-group transformation, summed over the trajectory, yields the free energy and therefore the thermodynamic densities.

The dimensionless free energy per bond $f = F/kN$ is thus obtained by summing the constants generated at each renormalization-group step,
\begin{equation}
f \, = \, \frac{1}{N} \ln \sum_{\{s_i\}} e^{-\beta {\cal H}} \, = \,
\sum_{n=0}^{\infty} \frac{G^{(n)}}{b^{dn}},
\end{equation}
where $N$ is the number of bonds in the initial unrenormalized system, the first sum is over all states of the system, the second
sum is over all renormalization-group steps $n$, $G^{(0)}$ is the constant from the first division at the beginning of the trajectory.
This sum quickly converges numerically.

A derivative of the free energy $f$ with respect to $J$ gives the energy density $<\delta (s_i,s_j)>$.  The thus calculated $q$-state Potts energy densities in $d=2$ and 3 are shown in Fig.3.  The latent heat discontinuities are shown with the dashed lines and are consistent with order of the phase transition yielded by the renormalization-group flows.  The correct $q_c=2$ is obtained for $d=3$.  In $d=2$, we need a first-order transition for $q=5$ to obtain $q_c=4$.  However, this is a near miss in the calculation, physically explained:  In the middle of a disordered island, all spin states contribute to the local multiplicity introduced above.  Thus, the subtraction $q-1$ is an oversubstraction.  In fact, when $q-0.25$ is used, the first-order transition with the latent heat is obtained, as seen in Fig.3.

\begin{figure}[ht!]
\centering
\includegraphics[scale=0.5]{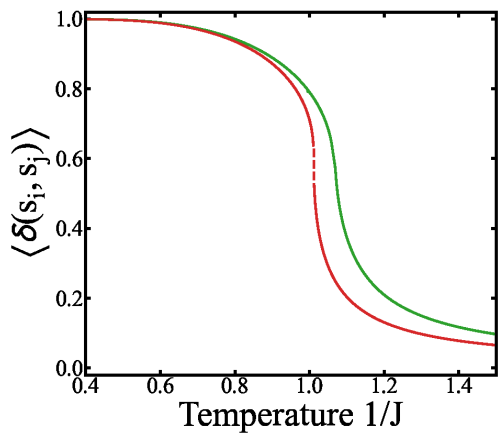}
\caption{Calculated energy density in $q=5$ and $d=2$.  The left curve uses the $q-0.25$ for the local disorder multiplicity.  The corrrect first-order phase transition is obtained (left curve) by simple, physically motivated adjustment.}
\end{figure}

\section{Conclusion}

The changeover from first-order to second-order phase transitions in q-state Potts models is obtained in spatial dimensions $d=2$ and 3 by a physically inspired simple adaptation of the simple Migdal-Kadanoff renormalization-group transformation.  The phase transition temperatures are dramatically improved by this physical adaptation.  The latent heats at the first-order phase transitions are calculated using the renormalization-group transformation. The inclusion of the local disorder state, which is the essence of our adaptation, can be used for numerical improvement and to take into account the possibility of a first-order phase transition.

\begin{acknowledgments}
Support by the Academy of Sciences of Turkey (T\"UBA) is gratefully acknowledged.
\end{acknowledgments}

\end{document}